\documentclass[%
reprint,
amsmath,amssymb,
aps,
]{revtex4-2}
\pdfoutput=1

\usepackage{graphicx}
\usepackage{dcolumn}
\usepackage{bm}
\usepackage{svg} 

\newcommand{\be}{\begin{equation}}
\newcommand{\ee}{\end{equation}}

\begin{document}
	
	\preprint{APS/123-QED}
	
	\title{Spectral Width of Maximum Deposition Eigenchannels in Diffusive Media}
	
	\author{Rohin E. McIntosh$^1$}
	\author{Arthur Goetschy$^2$}
	\author{Nicholas Bender$^3$}
	\author{Alexey~Yamilov$^4$}
	\author{Chia Wei Hsu$^5$}
	\author{Hasan Y\i lmaz$^6$}
	\author{Hui Cao$^{1,*}$}
	\affiliation{
		$^1$Department of Physics, Yale University, 06511 New Haven, CT, USA\\
		$^{2}$  Institut  Langevin, ESPCI  Paris, PSL  University,  CNRS, F-75005 Paris,  France \\
		$^{3}$  School of Applied and Engineering Physics, Cornell University, Ithaca, New York 14850, USA \\
		$^{4}$ Physics Department, Missouri University of Science \& Technology, Rolla, Missouri \\
		$^{5}$ Ming Hsieh Department of Electrical and Computer Engineering, University of Southern California, Los Angeles, California 90089, USA \\
		$^{6}$ Institute of Materials Science and Nanotechnology, National Nanotechnology Research Center (UNAM), Bilkent University, 06800 Ankara, Turkey \\
		$^*$hui.cao@yale.edu
	}
	
	
	\begin{abstract}
		
		The maximum deposition eigenchannel provides the largest possible power delivery to a target region inside a diffusive medium by optimizing the incident wavefront of a monochromatic beam. It originates from constructive interference of scattered waves, which is frequency sensitive. We investigate the spectral width of maximum deposition eigenchannels over a range of target depths using numerical simulations of a 2D diffusive system. Compared to tight focusing into the system, power deposition to an extended region is more sensitive to frequency detuning. The spectral width of enhanced delivery to a large target displays a rather weak, non-monotonic variation with target depth, in contrast to a sharp drop of focusing bandwidth with depth. While the maximum enhancement of power deposited within a diffusive system can exceed that of power transmitted through it, this comes at the cost of a narrower spectral width. We investigate the narrower deposition width in terms of the constructive interference of transmission eigenchannels within the target. We further observe that the spatial field distribution inside the target region decorrelates slower with spectral detuning than power decay of the maximum deposition eigenchannel. Additionally, absorption increases the spectral width of deposition eigenchannels, but the depth dependence remains qualitatively identical to that without absorption. These findings hold for any diffusive waves, including electromagnetic waves, acoustic waves, pressure waves, mesoscopic electrons, and cold atoms.
		
	\end{abstract}
	
	\maketitle
	
	\section{Introduction}
	
	Targeted delivery of light deep into random-scattering media has important applications in deep-tissue imaging \cite{Yu2015, Yoon2020}, optogenetics \cite{2011_Fenno, Yoon2015, 2017_Pegard_NC, Ruan2017}, laser microsurgery \cite{2004_Yanik, Yu2015}, photothermal therapy \cite{Pernot2007} and photopharmacology~\cite{velema2014photopharmacology, hull2018vivo}. Light penetration into a diffusive medium can be enhanced by tailoring the incident wavefront of a coherent beam~\cite{Mosk2012, Rotter2017, Cao2022}. Over the past decade, wavefront shaping techniques have been successfully applied to focusing light through or into multiple-scattering media~\cite{Vellekoop2007, Yaqoob2008, Vellekoop2008_2, Popoff2010, Hsieh2010, Xu2011, Judkewitz2013, Chaigne2014, Liu2015, Horstmeyer2015, Vellekoop2015, Katz2019, Boniface2019, Yang2019, Boniface2020}, as well as enhancing total transmission and power delivery inside~\cite{Vellekoop2008, Choi2011, Kim2012, Yu2013, Popoff2014, Gerardin2014, Davy2015, Hsu2017, Yilmaz2019, Bender2020, Choi2013, Cheng2014, Sarma2016, Jeong2018, Durand2019, Lee2023, Bender2022_2, Kim2020TimegatedIP, Cao2022EnhanceTD, Shaughnessy2024MultiregionLC}. It has been shown that the maximal enhancement of power delivered to an extended region inside a diffusive medium can exceed that of total transmission~\cite{Bender2022_2}.
	
	Deposition eigenchannels are introduced as an orthogonal set of input fields at a given frequency, the largest of which delivers the highest possible power to a designated region. The maximum deposition eigenchannel has a finite spectral width. As the input frequency is detuned from the frequency at which the channel is defined, the power within the target drops, eventually reaching the level of random wavefront illumination. The maximum deposition eigenchannel bandwidth and its dependence on the target depth and size are not yet understood. Frequency detuning also changes the spatial field distribution in the target. It is not known whether the field decorrelates faster or slower than the power decay. Furthermore, absorption is expected to impact the deposition channels, as it does to transmission channels~\cite{Liew2014, Liew2015, Sarma2015, Yamilov2016}.
	
	In this report, we conduct a numerical study on the spectral width of maximum deposition eigenchannels and their field decorrelation with frequency detuning. Compared to experimental studies, numerical simulations can provide the internal field distribution with sub-wavelength resolution. We find that the maximum deposition eigenchannel for an extended target deep inside a two-dimensional (2D) diffusive system has a narrower spectral width than the maximum transmission eigenchannel. With increasing depth, the deposition width first decreases gradually with depth, then increases slightly. In contrast, focusing light onto a wavelength-scale target inside the same system has a broader width, which decays monotonically with depth. By decomposing the maximum deposition eigenchannel into transmission eigenchannels, we show that its spectral width is dominated by the interference of constituent transmission eigenchannels, which substantially narrows the deposition width. The spatial field distribution in a large target decorrelates slower than the power decay with small frequency detuning. This property persists for all target depths and is unique to the largest deposition eigenchannels. Our numerical results are explained analytically using perturbation theory. Finally, we find that optical absorption increases the deposition width, but its depth dependence is similar to that without absorption.
	
	\begin{figure}
		\includegraphics[width=.48\textwidth]{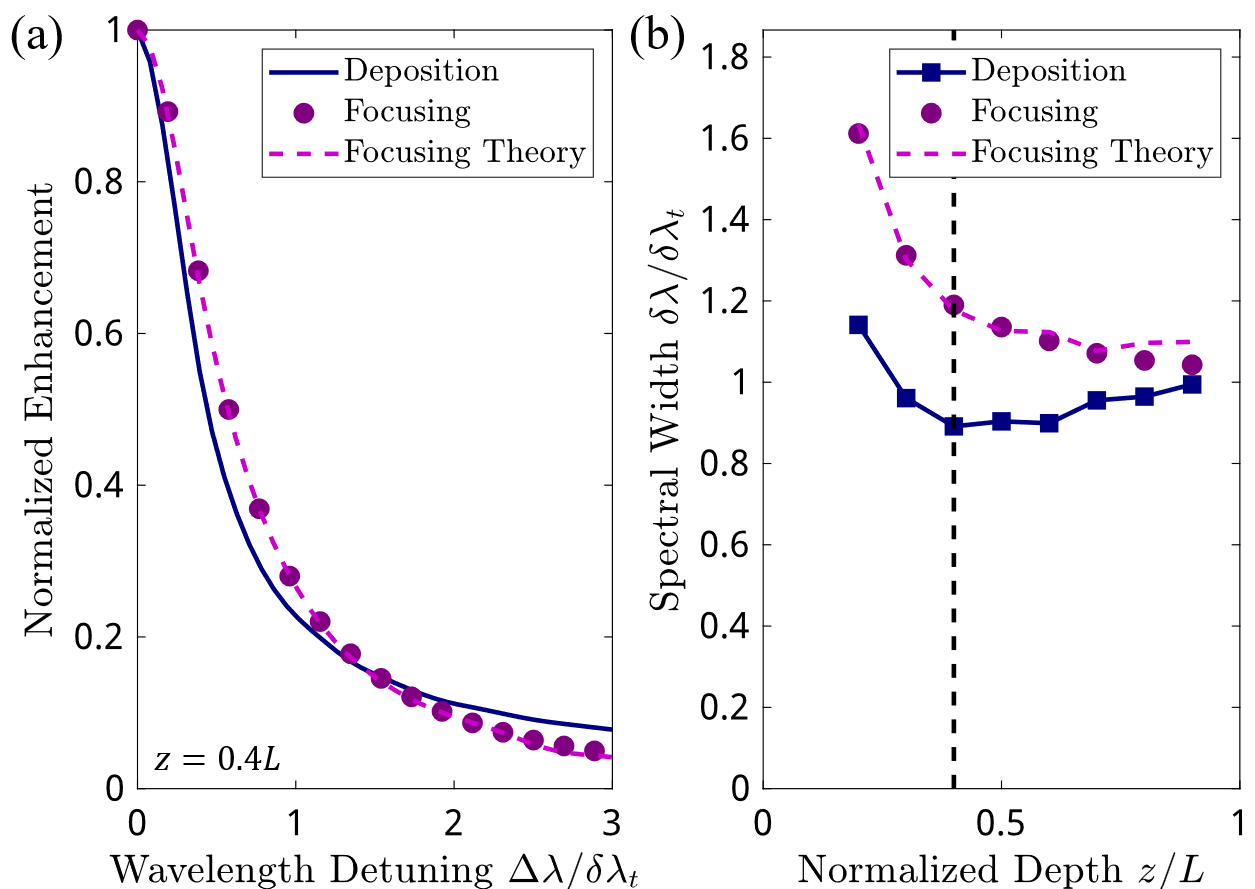}
		\caption{\label{Fig1} \textbf{Spectral bandwidth of the maximum deposition eigenchannel.} (a) Normalized power $\mathcal{C}^{(1)}(z,\Delta\lambda)$ deposited to a 10 \textmu m by 10 \textmu m region centered at depth $z = 0.4L$ decays with wavelength detuning $\Delta\lambda$ from $\lambda_0$ where the maximum deposition eigenchannel is defined (blue). It is compared to the decay of focused power to a wavelength-scale speckle $\mathcal{C}^{(\text{foc})}(z, \Delta\lambda)$ at $z=0.4L$ with wavelength detuning (purple), which agrees with analytic theory (purple dashed). Both curves are normalized such that they vanish at $\Delta \lambda \rightarrow \infty$. $\Delta\lambda$ is normalized by the spectral width of transmission $\delta\lambda_t$. (b) Spectral width in units of $\delta\lambda_t$ for the maximum deposition to a large target (blue squares) decreases and then increases slightly with depth $z$, normalized by system length $L$. For comparison, the spectral width of focusing to a wavelength-scale speckle (purple circles) decays monotonically with depth $z$, which agrees with analytic theory (purple dashed). The black dashed line marks the depth plotted in (a).}
	\end{figure}
	
	\section{Deposition Eigenchannels}
	
	Coherent wave transport through scattering media is described by the field transmission matrix, which maps a set of input spatial wavefronts to the transmitted field patterns. The largest singular vector of the transmission matrix gives the maximum transmission eigenchannel, which maximizes total transmission through the medium. Recently, the deposition matrix has been introduced to determine input wavefronts that maximize power delivery inside a scattering medium~\cite{Bender2022_2}. This matrix defines a linear mapping from the incident wavefront to the deposited field distribution over a target inside a scattering medium. The largest singular vector of the deposition matrix gives the deposition eigenchannel that delivers the maximal power to the target. The depth and dimensions of the target region can be chosen freely. The matrix elements are constructed by launching a complete set of orthogonal wavefronts and sampling the field inside the target. For a monochromatic light of wavelength $\lambda$,
	\be
	\mathcal{Z}_{mn}(\lambda)\equiv\sqrt{\epsilon({\bf r}_m)V/M}E_n({\bf r}_m,\lambda),
	\ee
	where $M$ is the total number of sampling points, ${\bf r}_m$ denotes the position of the $m$th sampling point, $V$ is the target volume, $E_n$ is the internal field resulting from the propagation of the $n$th input channel, $\epsilon(\bf r)$ is the spatially-varying dielectric constant, and $\lambda$ is the wavelength. The total power within the target for an incident wavefront $|\psi\rangle$ is given by $\langle\psi|\mathcal{Z}^\dag(\lambda)\mathcal{Z}(\lambda)|\psi\rangle$. Averaging over random incident wavefronts gives the mean power $\langle\zeta\rangle$ in the target. The complex conjugate of each row of the deposition matrix provides the input wavefront that focuses light on the corresponding sampling point in the target.
	
	To maximize the power delivered to the designated target, we perform singular value decomposition on the deposition matrix, $\mathcal{Z}(\lambda) = U\Sigma V^\dag$, which provides a set of right and left singular vectors as the columns of $V$ and $U$ respectively. The right singular vector $|v_1 (\lambda)\rangle$ corresponding to the largest singular value $\sqrt{\zeta_{1}(\lambda)}$ gives the spatial wavefront that maximizes the power in the target, the left singular vector $|u_1 (\lambda)\rangle$ is proportional to the resulting field distribution in the target, and $\zeta_{1}(\lambda)$ is the delivered power.
	
	The deposition matrix $\mathcal{Z}(\lambda)$ and its eigenchannels change with wavelength $\lambda$. As $\lambda$ shifts away from $\lambda_0$ where the maximum deposition eigenchannel is defined, the power delivered by the same input wavefront to the target decays. Namely, the input wavefront is fixed to $|v_1(\lambda_0)\rangle$, given by the maximum deposition eigenchannel at $\lambda_0$, and the power delivered to the target at $\lambda$ is given by,
 \be\zeta_{1}(\Delta\lambda) = \langle v_1(\lambda_0) |\mathcal{Z}^\dag(\lambda) \mathcal{Z}(\lambda) |v_1(\lambda_0)\rangle,
 \ee
 where $\Delta\lambda = |\lambda - \lambda_0|$ is the wavelength detuning. For large $\Delta\lambda$, $\zeta_{1}(\Delta\lambda)$ approaches the mean power $\langle\zeta(\lambda)\rangle$ for random input wavefronts. We define the deposition spectral width $\delta\lambda$ as the full-width-at-half-maximum (FWHM) of $\zeta_{1}(\Delta\lambda)$.
	
	Previous works have measured deposition eigenchannels experimentally by imaging light inside two-dimensional disordered waveguides from the third dimension~\cite{Bender2022_2, McIntosh2024}. These waveguides are fabricated in silicon-on-insulator wafers, and the disordered region consists of a random array of air holes. While these experiments convincingly demonstrate the enhancement of power delivery, there are a number of limitations. Primarily, field measurements inside the disorder waveguides are performed indirectly by imaging light scattered out of the plane. These measurements are limited by the spatial resolution of the detection and cannot resolve wavelength-scale speckles in the sample. The out-of-plane scattering inevitably induces loss, and it is impossible to switch off this loss to investigate the effect of absorption on deposition eigenchannels. In addition, only a handful of disorder realizations can be studied experimentally.
	
	Therefore, to directly study the spectral width of maximum deposition eigenchannels over many realizations, we numerically simulate light transport in a 2D disordered waveguide with the same physical parameters as in previous experiments. The waveguide supports $N=56$ propagating modes within a spectral range of $\lambda = 1502.25-1523.25$ nm. The disordered region has length $L=50$ \textmu m and width $W=15$ \textmu m. The transport mean free path is $\ell_t=3.3$ \textmu m. Thus light transport is diffusive, with $\ell_t\ll L$. To study the effects of absorption, we add an imaginary part to the dielectric constant inside the waveguide. The diffusive absorption length is $\xi = 28$~\textmu m. Our simulations are conducted using KWANT, an open-source Python package for coherent wave transport simulations~\cite{Groth2014}. We simulate identical disordered structures with and without absorption. The wavelength is tuned with a step size of $d\lambda = 0.25$ nm. All results shown represent an average of 100 independent disorder realizations.
	
	To calculate deposition matrices, we launch light in each of the $N$ propagating waveguide modes, and the respective fields within the target region are calculated. This creates a linear mapping from the input basis of propagating waveguide modes to the spatial basis of field distributions in the target region. The maximum deposition eigenchannels are calculated from the simulated matrices using singular value decomposition. For deposition, target regions are 10 \textmu m by 10 \textmu m squares centered on the target depth $z$. For comparison, we calculate the transmission matrix and its eigenchannels~\cite{Bender2020}. Transmission matrices map the incident fields to the transmitted fields, both in the waveguide mode basis. In all results, we normalize the spectral detuning $\Delta\lambda$ by the FWHM of maximum transmission eigenchannel, $\delta\lambda_{t} = 1.34$ nm.
	
	\section{Spectral Width of the Maximum Deposition Eigenchannel}
	
	We consider an extended target of dimension $10$ \textmu m by $10$ \textmu m, containing roughly $\sim1700$ speckles at a wavelength of around $1575$ nm. Figure~\ref{Fig1}a shows the optical power delivered at wavelength $\lambda =\lambda_0 + \Delta\lambda$ by the maximum deposition eigenchannel at $\lambda_0 = 1575.6$ nm,  to a target centered at $z=0.4L$ (blue line). The power is ensemble-averaged and normalized as,
 \be
\mathcal{C}^{(1)}(z,\Delta\lambda)=\frac{\langle\zeta_{1}(z, \Delta\lambda) \rangle - \langle\zeta(z)\rangle}
{\langle\zeta_{1}(z,0) \rangle - \langle\zeta(z)\rangle}.
\ee
The delivered power $\langle\zeta_{1}( z, \Delta\lambda) \rangle$ decays monotonically with wavelength detuning, approaching the value of a random wavefront $\langle\zeta (z) \rangle$. At large spectral detuning, the power decay slows down, exhibiting a long tail. This is consistent with the behavior of maximum transmission eigenchannel, which is attributed to long-range mesoscopic correlations~\cite{Hsu2015}. We note that the analytical computation of $\langle\zeta_{1}(z, \Delta\lambda) \rangle$ is difficult. So far, it has been computed for $\Delta\lambda=0$ only~\cite{Bender2022}, and the result already shows a complicated non-linear dependence on the long-range mesoscopic correlation $C_2 (z,\Delta\lambda=0)$ (see below for a precise definition of $C_2$).

For comparison, we also simulate light focusing on a single speckle grain at the same depth $z$. Representing this speckle grain by the state $|m\rangle$, the input wavefront $|\psi_{\text{foc}} (\lambda_0)\rangle $ is obtained from the conjugate of the $m^{th}$ row of the deposition matrix, $|\psi_{\text{foc}} (\lambda_0)\rangle = \mathcal{N}(\lambda_0) \mathcal{Z}(\lambda_0)^\dagger |m\rangle $, with $\mathcal{N}(\lambda_0) = \langle m |\mathcal{Z}(\lambda_0) \mathcal{Z}(\lambda_0)^\dagger | m \rangle^{-1/2}$. The normalized power at the focus at wavelength $\lambda=\lambda_0 + \Delta\lambda$ is defined by
 \be
 \mathcal{C}^{(\text{foc})}(z, \Delta\lambda) = \frac{\left<
\vert
\langle m |\mathcal{Z}(\lambda) 
|\psi_{\text{foc}} (\lambda_0)\rangle
\vert^2\right>
 }
 {
\left< \vert
\langle m |\mathcal{Z}(\lambda_0) 
|\psi_{\text{foc}} (\lambda_0)\rangle
\vert^2\right>
 },
\ee 
 which gives $ \mathcal{C}^{(\text{foc})}(\Delta\lambda) \simeq \left<|\langle\psi_{\text{foc}}(\lambda)|\psi_{\text{foc}}(\lambda_0)\rangle|^2\right>$, for $\mathcal{N}(\lambda) \simeq \mathcal{N}(\lambda_0) \simeq \left<\mathcal{N}(\lambda_0) \right>$. With the input wavefront fixed, the power at the focus decreases when the wavelength is detuned from $\lambda_0$. As shown in Fig.~\ref{Fig1}a, at small detuning, focusing decays slower than deposition to an extended target, but at large detuning, it becomes faster. This is because focusing is determined primarily by the short-range correlation $C_1(z, \Delta \lambda)$~\cite{Beijnum2011}. Our simulation results for focusing are well described by the following theoretical expression derived in Supplementary Material (dashed line in Figure~\ref{Fig1}a),
 \begin{align}
\mathcal{C}^{(\text{foc})}(z, \Delta\lambda) &=  C_1(z,\Delta\lambda) 
\nonumber
\\
&+ \left[1 - C_1(z,\Delta\lambda)\right]
\left[C_2(z,\Delta\lambda) + \frac{1}{N}\right].
\label{EqFocusingTheory}
\end{align}
Here, $C_1(z,\Delta\lambda)$ and $C_2(z,\Delta\lambda)$ are the short- and long-range contributions to the spectral correlation function of the total intensity, $\mathcal{C}^{(T)}(z,\Delta\lambda)=\left<I(z,\lambda_0) I(z,\lambda) \right>/\left<I(z,\lambda_0) \right>\left< I(z,\lambda) \right> -1$, 
    where $I(z,\lambda)=\sum_b \vert\mathcal{Z}_{ba}(\lambda)\vert^2$. 
     The analytical expressions of $C_1(z,\Delta\lambda)$ and $C_2(z,\Delta\lambda)$ have been derived in Ref.~\cite{McIntosh2024}. In the diffusive limit $L\gg \ell_t$ and at $\Delta\lambda=0$, $C_1(z,0)=1$ and  $C_2(z,0)=(4z/\pi N\ell_t)(1-2z/3L)$. The theoretical prediction shown in Fig.~\ref{Fig1} corresponds to Eq.~\eqref{EqFocusingTheory}, where $C_1(z,\Delta\lambda)$ and $C_2(z,\Delta\lambda)$ have been computed numerically, following a procedure detailed in Supplementary Material. Interestingly, we note that the power at the focus acquires a dependence on the long-range component $C_2(z,\Delta\lambda)$ for $\Delta\lambda\neq 0$, which is smaller than $C_1(z,\Delta\lambda)$ but not negligible in the results presented in Fig.~\ref{Fig1}. 
	
	To find how the spectral width of the maximum deposition eigenchannel varies with depth $z$, we move the $10$ \textmu m by $10$ \textmu m target region throughout the diffusive waveguide, and compute $\mathcal{C}^{(1)}(z,\Delta\lambda)$ for different $z$. In Fig.~\ref{Fig1}b, the spectral width $\delta \lambda$, given by the FWHM of $\mathcal{C}^{(1)}(z,\Delta\lambda)$, is normalized by $\delta\lambda_t$ and plotted versus the normalized depth $z/L$. As the target depth increases, the deposition width first drops gradually and then rises near $z=L$. The spectral width of the maximum transmission eigenchannel is greater than that of the maximum deposition channel in most of the interior of the waveguide. Figure~\ref{Fig1}b also shows the spectral width for focusing to a single speckle as a function of depth. It is given by the FWHM of $\mathcal{C}^{(\text{foc})}(z, \Delta\lambda)$. The focusing width drops monotonically with depth, in contrast to the weak, non-monotonic variation of the width for deposition to an extended target, which is minimized inside the waveguide. We further show that our analytic model~\eqref{EqFocusingTheory} predicts the focusing spectral width for any depth, confirming that focusing on a single speckle inside the system is primarily dictated by short-range correlations.  
  	
	\begin{figure}
		\includegraphics[width=.48\textwidth]{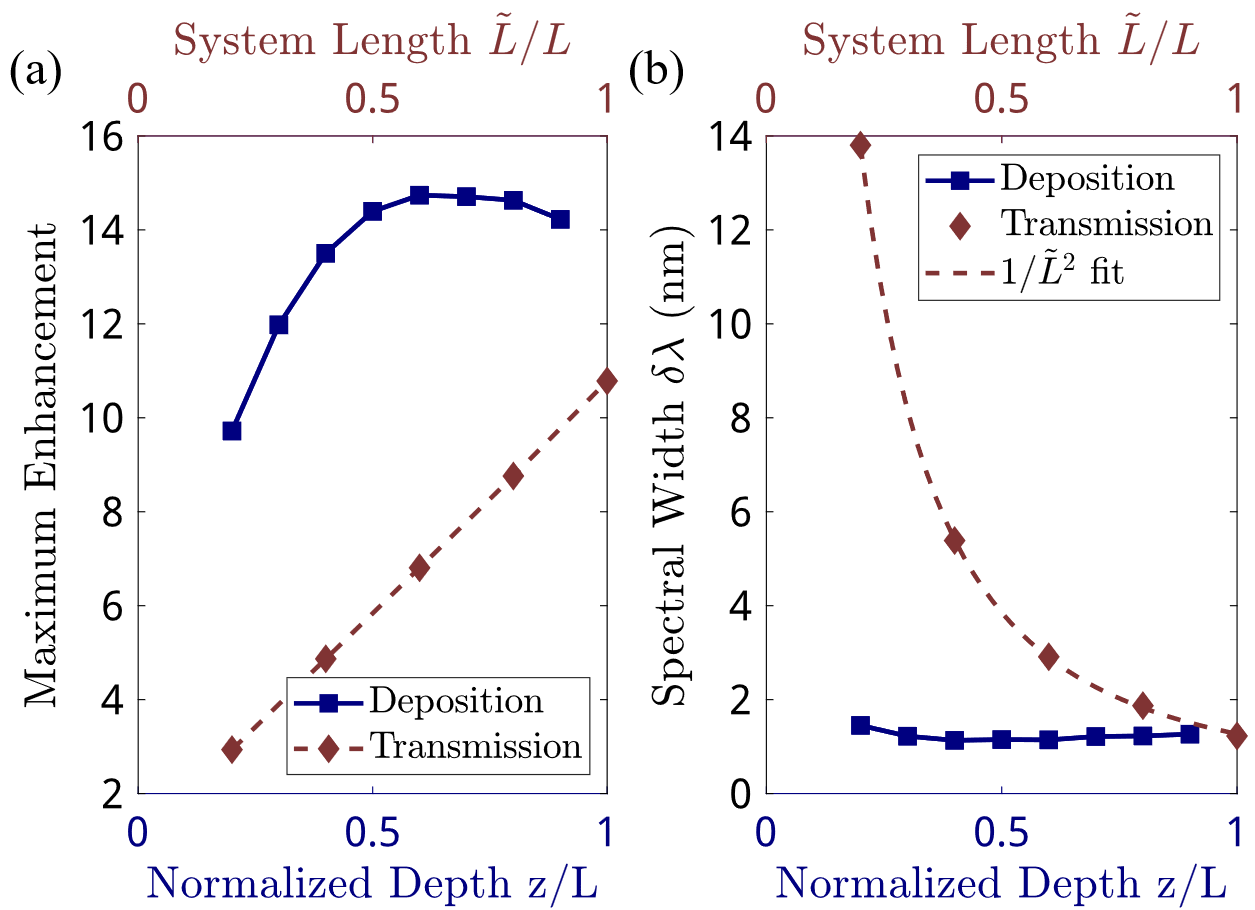}
		\caption{\label{Fig2} \textbf{Comparison between maximum deposition and transmission eigenchannels.} (a) Maximum deposition enhancement $\zeta_{1}/\langle\zeta\rangle$ for a 10 \textmu m by 10 \textmu m target centered at depth $z/L$ (blue squares) of a diffusive waveguide of length $L$ increases and then decreases with depth. It contrasts the linear increase of the maximum transmission enhancement through a diffusive waveguide of length $\tilde{L}=z$ (red diamonds). The deposition enhancement inside the waveguide exceeds transmission enhancement, both compared to random input wavefronts. (b) The spectral width of the power decay for the maximum deposition eigenchannel (blue squares) is smaller than that of the maximum transmission eigenchannel (red diamonds) at all depths. The latter scales as $1/\tilde{L}^2$, as confirmed by the curve fitting (red dashed). The deposition width shows a comparatively small change with the target depth.}
	\end{figure}
	
	\section{Comparison of Maximum Deposition and Maximum Transmission Eigenchannels}

 We wish to highlight the difference between the maximum deposition and transmission eigenchannels by isolating the effects of forward and backpropagating waves entering the target. To do this, we compare the maximum deposition eigenchannel at depth $z$ in a diffusive waveguide of length $L$ to the maximum transmission eigenchannel of a shortened waveguide of length $\tilde{L} = z$, where the section from $L-z$ to $L$ of the waveguide has been removed.
 
	In Figure~\ref{Fig2}a, we plot the maximum transmission enhancement vs. $\tilde{L}/L$ alongside the maximum deposition enhancement for a $10$ \textmu m by $10$ \textmu m target centered at $z/L$. The transmission enhancement scales as $\tilde{L}/\ell_t$~\cite{Pnini1989}, while the deposition enhancement is maximized at $z/L\simeq 3/4$, exceeding the transmission enhancement at $z=L$~\cite{Bender2022_2}. Notably, the deposition enhancement at all target depths $z<L$ exceeds the transmission enhancement for the waveguide of length $\tilde{L}=z$. The two become equal at $\tilde{L}= z = L$.
 
  The higher deposition enhancement throughout the waveguide is explained by the constructive interference of waves entering the target from all directions. In the case of transmission, only the waves propagating to the exit facet contribute to the transmitted light. For internal deposition, however, scattered waves can enter the target from all directions. In the diffusive waveguide, for example, waves may pass through the target and then return after multiple scattering. As a result, power in a target at depth $z$ is enhanced by constructive interference of waves coming from both sections before and after the target region, while transmission enhancement through the waveguide of length $\tilde{L} = z$ only has contributions from the first section. The saturation of the deposition enhancement is a finite-size effect. As the target nears the exit facet, diffusive waves propagating past the target have a higher chance of leaving the sample instead of returning back to the target. This reduces the contribution of backpropagating waves to the power enhancement in the target, thereby lowering the enhancement.
	
	The larger enhancement factor for deposition, however, corresponds to a narrower spectral range of enhancement. Figure~\ref{Fig2}b shows that the spectral width of the maximum transmission eigenchannel decreases as $1/\tilde{L}^2$ with the diffusive system length $\tilde{L}$~\cite{Yamilov2005}, while the transmission enhancement increases with $L$. For comparison, the spectral width of the maximum deposition channel for a 10 \textmu m by 10 \textmu m target centered at depth $z = \tilde{L}$ is narrower than that of maximum transmission through $\tilde{L}$. This is attributed to diffusive waves returning to $z$ from the section between $z$ and $L$. Their interference with the waves reaching $z$ from $z<L$ enhances not only the power at $z$ but also spectral sensitivity. For $z \ll L$, the depth dependence of the spectral width for the maximum deposition eigenchannel decays as $1/z^2$, similar to the scaling of transmission. With increasing $z$, the decay is replaced by a nearly invariant width as a result of the finite length of the diffusive waveguide.
	
	\section{Decomposition into Transmission Eigenchannels}
	
	\begin{figure*}
		\includegraphics[width=.7\textwidth]{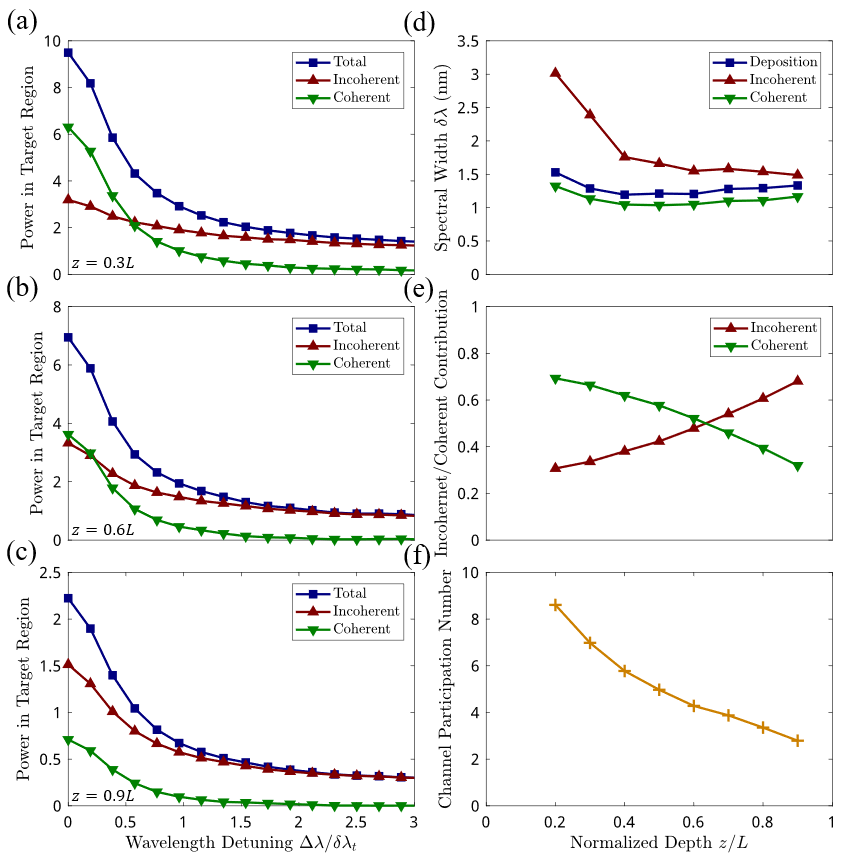}
		\caption{\label{Fig3} \textbf{Decomposition of the maximum deposition eigenchannel by transmission eigenchannels.} (a-c) Optical power $\zeta_{1}(\Delta\lambda)$ deposited by the largest deposition eigenchannel defined at $\lambda_0$ at three target depths $z/L = 0.3,\ 0.6,\ 0.9$ (blue) is decomposed into the sum of incoherent (red) and coherent (green) contributions of transmission eigenchannels at the detuned wavelength $\lambda$. (d) The spectral width of the total power (blue squares) is compared to that of incoherent contribution (red upward triangles) and of coherent contribution (green downward triangles). The coherent contribution of transmission eigenchannels has a narrower width, which dominates the maximum deposition channel width. (e) The percentage of coherent contribution $P_c/P_1$ to the power of maximum deposition channel decays with depth (green downward triangles), while the percentage of incoherent contribution $P_i/P_1$ increases at $\lambda = \lambda_0$. (f) Participation number of transmission eigenchannels to maximum deposition $M_{e}(z,0)$ decays with target depth $z/L$ at $\Delta \lambda = 0$.}
	\end{figure*}

The spectral width of the maximum deposition eigenchannel exhibits a peculiar dependence on the target depth, attaining a minimum inside the waveguide. This contrasts the non-monotonic depth dependence of the deposition enhancement, which is maximized inside the waveguide. However, a precise explanation for the depth dependence is difficult due to the lack of an analytic model for $\langle\zeta_{1}(z, \Delta\lambda) \rangle$. To probe the effect of wave interference on the depth dependence of the deposition spectral width, we decompose the maximum deposition eigenchannel at wavelength $\lambda_0$ into a superposition of transmission eigenchannels at wavelength $\lambda$ in a manner similar to the approach of Ref.~\cite{Bender2022_2}. Our results show that the contribution to the delivered power by transmission eigenchannels interfering inside the target has a similar depth dependence to the spectral width of the maximum deposition eigenchannel.
 
 The decomposition of an input wavefront at $\lambda$, corresponding to the largest deposition eigenchannel defined at $\lambda_0$, reads $|v_1(z,\lambda_0)\rangle = \sum_{\beta=1}^N d_\beta (z, \lambda_0, \lambda) \, |v^t_\beta(\lambda)\rangle$, where $|v^t_\beta(\lambda)\rangle$ are the orthogonal right singular states of the transmission matrix, and $d_\beta$ are decomposition coefficients that depend on both $\lambda$ and $\lambda_0$, as well as the target depth $z$. Due to linear mapping from the incident fields to fields in the target region, the decomposition of the internal field distribution by transmission channels has the same coefficients: $E_1({\bf r}, \lambda_0) = \sum_{\beta=1}^N d_{\beta}(z, \lambda_0, \lambda) \, E^t_\beta({\bf r}, \lambda)$, where $E^t_\beta({\bf r}, \lambda)$ denotes the field distribution of the $\beta$th transmission eigenchannel in the target.
	
	Since transmission eigenchannels are not guaranteed to be orthogonal inside the scattering system, they can interfere in the target region.
	The optical power of the maximum deposition channel $P_1(z, \Delta\lambda)= \langle\zeta_{1}(z, \Delta\lambda) \rangle$ in the 10 \textmu m by 10 \textmu m target region consists of two terms,
    \be
	\begin{aligned}
		&P_1(z, \Delta\lambda) = P_{i}(z, \Delta\lambda) + P_{c}(z, \Delta\lambda) \\
		&= \left< \sum_{\beta=1}^N |d_{\beta}|^2 P_\beta(z,\lambda) \right>
  + \left<\sum_\beta \sum_{\beta'\neq\beta} d_{\beta} d^*_{\beta'} P_{\beta \beta'}(z, \lambda)
\right>.
	\end{aligned}
    \ee
	where $P_\beta(z,\lambda) = \int |E^t_\beta({\bf r}, \lambda)|^2 d{\bf r}$ is the power of $\beta$th transmission eigenchannel in the target region centered at depth $z$, and $P_{\beta \beta'}(z,\lambda) = \int E^t_\beta({\bf r}, \lambda) \, E^{t*}_{\beta'}({\bf r}, \lambda) d{\bf r}$ is the interference between the $\beta$th and $\beta'$th transmission channels within the target. Hence, the first term $P_i (z, \Delta\lambda)$ is an incoherent sum of transmission eigenchannels in the target, and the second term $P_c (z, \Delta\lambda)$ represents the coherent contribution of transmission eigenchannels to the power inside the target. The coherent contribution depends on the relative phase between transmission eigenchannels. Without the coherent term, the incoherent contribution alone cannot enhance the power in a large target more than the highest possible enhancement by a single transmission eigenchannel.
 
 Figure~\ref{Fig3}(a-c) illustrates the decay of the ensemble-averaged powers $P_i (z, \Delta\lambda)$ and $P_c (z, \Delta\lambda)$ with spectral detuning $\Delta\lambda$ at three depths: $z/L = 0.3,\ 0.6,\ 0.9$. While the total power $P_1 (z, \Delta\lambda)$ and the incoherent contribution $P_i (z, \Delta\lambda)$ both decay to $\langle\zeta\rangle$ at large spectral detuning, the coherent contribution $P_c (z, \Delta\lambda)$ decays to zero as channel interference vanishes at $z=L$.
		Figure~\ref{Fig3}d shows that the FWHM of the incoherent contribution $P_i$ is far greater than that of coherent contribution $P_c$. This is because the latter relies on wave interference which is inherently narrowband. Most notably, the coherent contribution is dominant in determining the deposition spectral width and its depth dependence.
	
	To quantify the percentage of incoherent and coherent contributions to the maximum deposition eigenchannel, we calculate $P_i(z, \Delta\lambda)/P_1(z, \Delta\lambda)$ and $P_c(z, \Delta\lambda)/P_1(z, \Delta\lambda)$ at varying depth $z$ and $\Delta\lambda=0$. As shown in Figure~\ref{Fig3}e, at $\lambda = \lambda_0$, $P_i(z, 0)/P_1(z, 0)$ increases with depth, while $P_c(z, 0)/P_1(z, 0)$ decreases continuously to zero at $z=L$. The reduction in the coherent contribution is attributed to a smaller number of constituent transmission eigenchannels at larger $z$, where low-transmission channels almost die out. This explanation is confirmed by the depth dependence of the participation number of transmission channels, defined as,
 \be
 M_{e}(z, \Delta \lambda) = \left<\frac{\left[\sum_{\beta=1}^N |d_{\beta}(z, \lambda_0, \lambda)|^2\right]^2}{N\sum_{\beta=1}^N |d_{\beta}(z, \lambda_0, \lambda)|^4}\right>.
 \ee
 Figure~\ref{Fig3}f shows that $M_{e}(z,0)$ decreases monotonically with $z$, indicating a decrease in the effective number of participating transmission channels as $z\rightarrow L$.
	
	\section{Field Decorrelation with Spectral Detuning}
	
	\begin{figure*}
		\includegraphics[width=.96\textwidth]{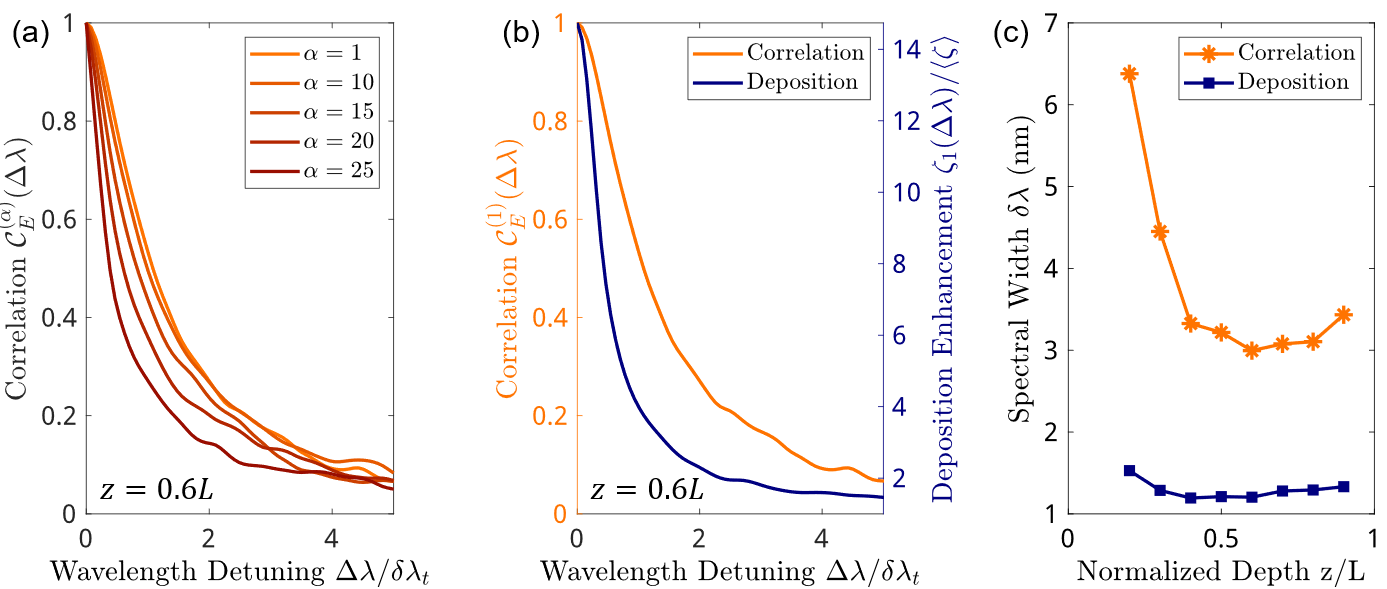}
		\caption{\textbf{Spectral field decorrelation of deposition eigenchannels.} (a) Spectral field correlation $\mathcal{C}_E^{(1)}(z,\Delta \lambda)$ of deposition eigenchannels $\alpha = 1,\ 10,\ 15,\ 20,\ 25$ show faster decorelation for lower deposition eigenvalue (larger $\alpha$). The target is a 10 \textmu m by 10 \textmu m box centered at depth $z = 0.6L$. (b) Spectral field correlation of the maximum deposition eigenchannel $\mathcal{C}_E^{(1)}(z,\Delta \lambda)$ decays with wavelength detuning $\Delta \lambda$ (orange, left axis) slower than the power enhancement by the the maximum deposition eigenchannel $\zeta_{1}(\Delta\lambda)/\langle\zeta\rangle$ (blue, right axis). The spectral detuning is normalized by the spectral width of transmission $\delta\lambda_t$. (c) Spectral width of the field correlation $\mathcal{C}_E^{(1)}(z,\Delta \lambda)$ (orange asterisks) is larger than that of power decay (blue squares) at all target depth. \label{Fig4}}
	\end{figure*}
	
	In addition to the power decay, spectral detuning changes the spatial field distribution in the target region for the maximum deposition eigenchannel. The spectral correlation function of the field distribution for input equal to the $\alpha$-th deposition channel at $\lambda_0$, $|v_\alpha(\lambda_0)\rangle$,
 is defined as,
	\be
	\mathcal{C}_E^{(\alpha)}(z,\Delta \lambda)=
 \bigg|\left< \frac{\langle v_\alpha (\lambda_0) \vert \mathcal{Z}^\dag(\lambda) \mathcal{Z}(\lambda_0) \vert v_\alpha(\lambda_0) \rangle}
 {\vert \vert \mathcal{Z}(\lambda)\vert v_\alpha(\lambda_0) \rangle\vert \vert \,  \vert \vert\mathcal{Z}(\lambda_0)\vert v_\alpha \rangle\vert \vert}\right>\bigg|.
	\label{EqDefFieldCorr}
	\ee
	We define the spectral width of the field correlation function as the FWHM of $\mathcal{C}_E^{(\alpha)}(z,\Delta \lambda)$.

	Figure~\ref{Fig4}a shows $\mathcal{C}_E^{(\alpha)}(z,\Delta \lambda)$ for high-deposition eigenchannels in a $10$ \textmu m by $10$ \textmu m target centered at depth $z/L=0.6$. Channels with larger eigenvalues (higher power in the target) have slower field decorrelation with wavelength detuning $\Delta \lambda$. This dependence can be understood as follows. When the input wavefront is set to that of the $\alpha$th deposition eigenchannel but its wavelength is slightly detuned, it predominantly excites the $\alpha$th eigenchannel along with a superposition of the remaining deposition eigenchannels in the diffusive waveguide. As its eigenvalue increases, the $\alpha$th eigenchannel becomes more dominant over the remaining eigenchannels, leading to a slower decay in correlation with frequency detuning. This behavior persists at all depths and is analogous to the larger angular memory effect for higher-transmission eigenchannels~\cite{Yilmaz2019memory}. It is also consistent with experimental observation of robustness against frequency detuning for high-transmission channels in a diffusive slab~\cite{Bosch2016}.
	
	Compared to power decay, field decorrelation of the largest deposition eigenchannel, $\mathcal{C}_E^{(1)}(z,\Delta \lambda)$, is slower at small spectral detuning $\Delta \lambda$. This is shown in Fig.~\ref{Fig4}b comparing the field decorrelation to the deposited power decay $\langle\zeta_1(z,\Delta \lambda)\rangle/\langle\zeta(z)\rangle$ as a function of $\Delta \lambda$. For large $\Delta \lambda$, the power decay becomes slower than the field decorrelation due to long-range spectral correlations~\cite{Hsu2015, McIntosh2024}. Figure~\ref{Fig4}c shows the dependence of the spectral width of the field correlation of the largest deposition eigenchannel, FWHM of $\mathcal{C}_E^{(1)}(z,\Delta \lambda)$, on the target depth $z$. At all depths, the field correlation spectral width exceeds that of the power enhancement. The former decreases rapidly for shallow depths before attaining a minimum near the center of the waveguide, where constructive interference between forward and backpropagating waves is strongest. 
 
	Our observation that the field decorrelation is slower than power decay at small wavelength detuning can be explained using a perturbation approach. For the input wavefront equal to the maximum deposition eigenchannel at $\lambda_0$, the field distribution inside the target is $|u_1\rangle = \mathcal{Z}(\lambda_0) \, |v_1\rangle$. The field variation induced by spectral detuning of same input wavefront is noted $ |\Delta u \rangle = \mathcal{Z}(\lambda_0) \, |v_1\rangle - \mathcal{Z}(\lambda_0 + \Delta \lambda) \,|v_1\rangle$, and is treated as a small perturbation of order $\epsilon \ll 1$ for small detuning. Up to a second-order expansion in $\epsilon$, the power change in the target region is 
	\be
		\frac{\langle\zeta_1(z,\Delta \lambda)\rangle} {\langle\zeta_1(z,0)\rangle} =\left<1 - \frac{2\Re[\langle u_1 \vert \Delta u_1 \rangle]} {\langle u_1 \vert u_1\rangle }  + \frac{\langle \Delta u_1 \vert \Delta u_1 \rangle} {\langle u_1\vert u_1\rangle}\right>,  
	\ee
	whereas the change in the spectral correlation function of field distribution can be written as
	\be
	\mathcal{C}_E^{(1)}(z,\Delta \lambda)  =\left<1  -  \frac{\langle \Delta u_1 \vert \Delta u_1 \rangle} {2 \langle u_1\vert u_1 \rangle} + \frac{\vert \langle u_1 \vert \Delta u_1 \rangle \vert^2} {2\langle u_1\vert u_1 \rangle^2}\right> \, .
	\ee
While the power decay has both first- and second-order corrections in $\epsilon$, the field decorrelation only has second-order corrections. Therefore, the power decay is faster than the field decorrelation for small spectral detuning. This calculation holds at all depths and explains the numerical results in Figure~\ref{Fig4}.
	
	\section{Effect of absorption}
	
	\begin{figure*}
		\includegraphics[width=.96\textwidth]{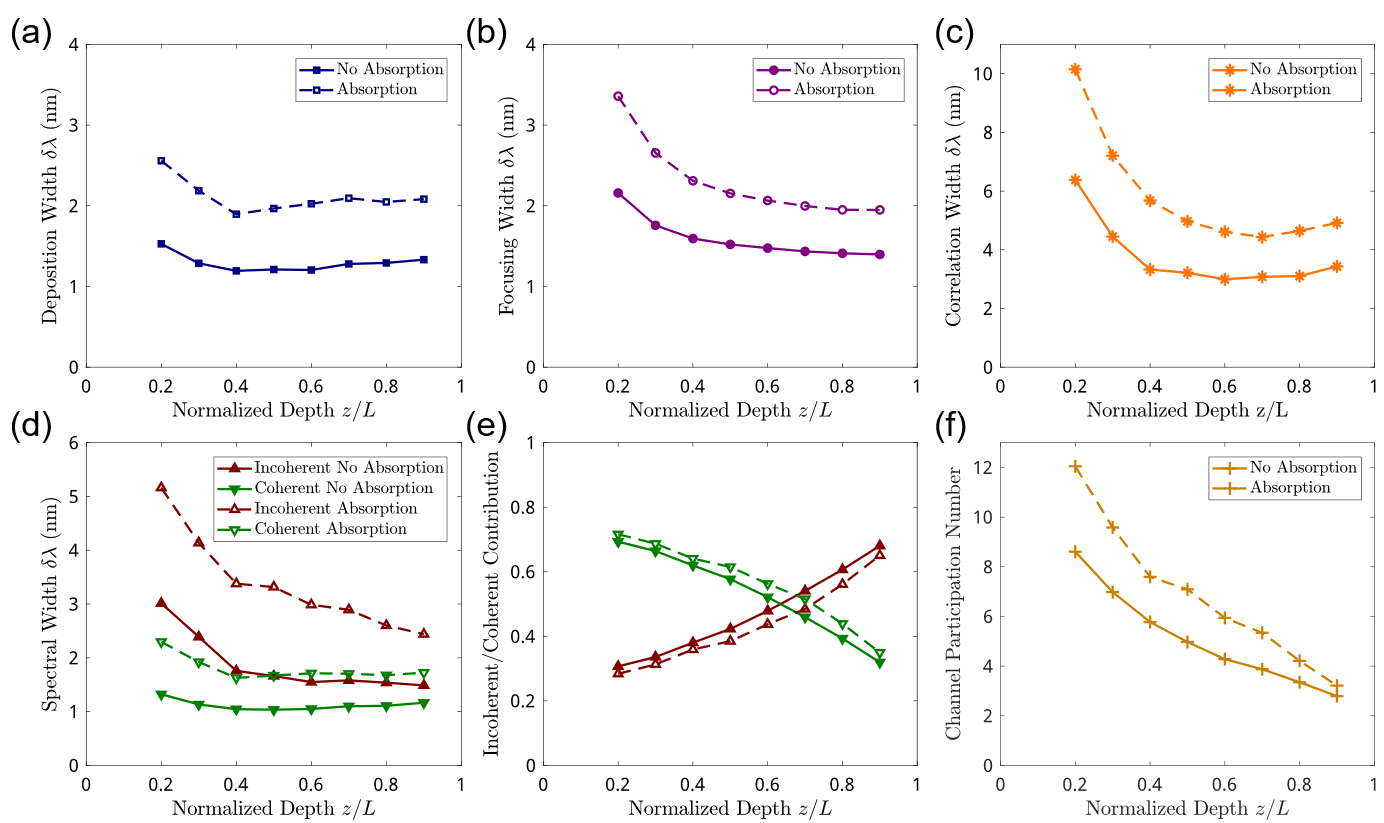}
		\caption{\textbf{Effect of absorption on maximum deposition eigenchannel.}  (a) Spectral width of the maximum deposition eigenchannel $\zeta_{1}(\Delta\lambda)$ (blue squares) with absorption (blue open squares) is larger than that without absorption (blue solid squares). The target is a 10 \textmu m by 10 \textmu m box centered at depth $z/L$. (b) The spectral width of focusing on a wavelength-scale speckle with absorption (purple open circles) exceeds that without absorption (purple solid circles) at all depths $z/L$. (c) The spectral width of the field correlation $\mathcal{C}_E^{(1)}(z,\Delta \lambda)$ in the 10 \textmu m by 10 \textmu m target with absorption (orange asterisks, dashed line) exceeds that without absorption (orange asterisks, solid line). (d) Spectral width of the incoherent contribution (red open upward triangles) and the coherent contribution (green open downward triangles) of transmission eigenchannels to the maximum deposition eigenchannel with absorption exceeds those without absorption (solid symbols). (e) Percentage of the incoherent (red upward triangles) contribution of transmission eigenchannels to the maximum deposition eigenchannel $P_i/P_1$ increases with absorption, while the coherent contribution $P_c/P_1$ decreases. (f) Participation number of transmission eigenchannels to the maximum deposition $M_{e}(z,0)$ with absorption (dashed line) is higher than that without absorption (solid line) at all depths $z/L$. \label{Fig5}}
	\end{figure*}
	
	It is known that absorption has a significant impact on coherent wave transport~\cite{Liew2014, Liew2015, Sarma2015, Yamilov2016}. Here we investigate how absorption affects the spectral width of maximum deposition eigenchannels. We introduce homogeneous absorption to the 2D diffusive waveguide by adding an imaginary part of the dielectric constant throughout the waveguide. The diffusive absorption length is $\xi = 28$ \textmu m. The target is still a $10$ \textmu m by $10$ \textmu m box centered at depth $z$. We also compare focusing light to a single speckle at the same depth.
	
	Figure~\ref{Fig5}a shows that absorption increases the spectral width of the maximum deposition eigenchannel at all depths. The width dependence on depth, however, remains qualitatively identical to that without absorption. The trade-off between power enhancement and spectral bandwidth persists in the presence of absorption. The spectral width of the maximum deposition eigenchannel is again minimized inside the waveguide. Therefore, absorption lowers the maximum power enhancement throughout the waveguide~\cite{Bender2022_2} while simultaneously increasing the spectral bandwidth.
 
    Figure~\ref{Fig5}b shows absorption also increases the spectral width for focusing light to a single speckle. However, the maximum enhancement of the focusing power remains unchanged by absorption. The increase in the spectral width results from the dominant contribution of the short-range correlation, whose spectral width increases with absorption.
    
    The focusing width becomes smaller than the deposition width for $z/L > 0.6$ in the presence of absorption. This is because absorption reduces the maximum power enhancement at $\Delta \lambda = 0$, and therefore power decay at large $\Delta \lambda$ (Figure~\ref{Fig1}a) plays an important role in determining the spectral width. At large $\Delta \lambda$, long-range spectral correlations become dominant, which are barely changed by absorption~\cite{Hsu2015, McIntosh2024}. This leads to an effective increase in spectral width. In contrast, the maximum focusing enhancement at $\Delta \lambda$ = 0 is unaffected by absorption, and the enhancement of the focusing bandwidth is primarily caused by the increase of the short-range spectral correlation width with absorption. 
	
	The introduction of absorption also increases the field correlation width in the target for the maximum deposition eigenchannel, FWHM of $\mathcal{C}_E^{(1)}(z,\Delta \lambda)$, as shown in Figure~\ref{Fig5}c. The depth dependence is again maintained with absorption. The width of $\mathcal{C}_E^{(1)}(z,\Delta \lambda)$ remains higher than that of the deposited power decay at all depths. $\mathcal{C}_E^{(1)}(z,\Delta \lambda)$ is determined by a combination of both short- and long-range correlations, whose spectral widths are both broadened by absorption. This results in an overall increase in the field correlation width with absorption.
	
	We further show that absorption increases the spectral width of incoherent and coherent contributions of transmission eigenchannels to the maximum deposition eigenchannel and preserves their depth dependence (Figure~\ref{Fig5}d). These results illustrate that the width increase of the incoherent contribution is more significant than that of the coherent contribution. This trend causes a notable increase in the deposition width as $z$ approaches $L$, where the coherent contribution vanishes. However, absorption slightly increases the magnitude of coherent contribution, $P_c(z, 0)/P_1(z,0)$, as shown in Figure~\ref{Fig5}e. The crossing between the coherent and incoherent contributions is at a larger depth than that without absorption. To explain this change, we calculate the number of participating transmission eigenchannels $M_e(z,0)$ to the maximum deposition eigenchannel with and without absorption. Figure~\ref{Fig5}f shows that absorption increases $M_e(z,0)$, thereby raising the magnitude of the coherent contribution. Although absorption decreases the power delivered by the maximum deposition eigenchannel, the number of contributing transmission eigenchannels increases at all target depths. This is because absorption causes a stronger attenuation of higher-transmission eigenchannels~\cite{Yamilov2016}, making a larger number of moderately transmitting channels contribute more to the largest deposition eigenchannel.
	
	\section{Discussion and Conclusion}

    In this work, we present a numerical study on the spectral bandwidth of the maximum deposition eigenchannel to an extended target deep inside a 2D diffusive waveguide. Compared to the maximum transmission channel, the maximum deposition eigenchannel is more sensitive to spectral detuning, resulting in a faster decay of deposited power in the target. The higher spectral sensitivity is a consequence of back-propagating waves interfering constructively with forward-propagating waves to maximize power in the target. In contrast to the monotonic decrease of transmission bandwidth with system length, the deposition bandwidth displays a non-monotonic dependence on the target depth, with the minimum inside the system. This is attributed to the finite length of the diffusive system.  We further show that the spectral decorrelation of the field distribution in the target region is slower than the power decay and provide an explanation with a perturbation approach. Finally, we find that optical absorption increases the spectral width for the maximum deposition eigenchannel at all target depths. 
    
    We stress the difference between delivering power to a target of size much larger than the wavelength and focusing light to a wavelength-scale speckle inside the diffusive waveguide. The spectral bandwidth of focusing is significantly larger and displays a monotonic decrease with increasing depth. This is due to the dominant contribution of short-range correlations to wavelength-scale focusing. For power delivery to a large target, long-range correlations become dominant, leading to distinct characteristics.
 
	Our results can be generalized to deposition eigenchannels in 3D diffusive systems. However, it is much more difficult to have complete control of the input wavefront in a 3D system compared to a 2D waveguide, as is done in this study. Therefore, it is important in future work to study the effect of incomplete channel control on power deposition inside 3D scattering media. It is worth noting that our simulation result in Fig.~\ref{Fig4}c conflicts with experimental results for high-transmission channels through a disordered slab measured by digital phase conjugation~\cite{Bosch2016}, which showed the spectral field correlation width of a high-transmission channel narrower than the transmission decay width. Although the conditions for this experiment were significantly different from those of our simulation, this discrepancy warrants further investigation. The spectral width for maximum power deposition may also depend on the overall size and shape of a scattering system~\cite{Rates2023}. It is also worth exploring deposition eigenchannels in systems with weak scattering, correlated disorder~\cite{uppu2021spatially}, or in the localized regime~\cite{pena2014single}. Finally, we point out that the current work applies to the power deposition of general diffusive waves, including microwaves, pressure waves, acoustics, or mesoscopic electrons. Furthermore, understanding the spectral properties of monochromatic deposition eigenchannels is important to broadband power delivery into diffusive media~\cite{McIntosh2024}. \\

{\bf Acknowledgement}

We thank Allard Mosk for valuable discussions. This work is supported partly by the US National Science Foundation (NSF) under Grants Nos. DMR-1905465 and DMR-1905442, and by the US Office of Naval Research (ONR) under Grant No. N00014-221-1-2026, and by the French Government under the program Investissements d’Avenir. 

\newpage

 {\bf Supplementary Material}
 \label{Supplementary Material}
 
\section*{Focus power at wavelength $ \lambda= \lambda_0 +\Delta \lambda$ }

The focus power is defined as
\be
P_{\text{foc}}(\Delta \lambda) = \left<
\vert
\langle m |\mathcal{Z}(\lambda) 
|\psi_{\text{foc}} (\lambda_0)\rangle
\vert^2\right>,
\ee
where $|m\rangle = E(r_m, \lambda)$ is the field at focus ($m$th sampling point), and $|\psi_{\text{foc}} (\lambda_0)\rangle = \mathcal{N}(\lambda_0) \mathcal{Z}(\lambda_0)^\dagger |m\rangle $ is the input wavefront that maximizes the focusing power for $\Delta \lambda=0$ .  The normalization coefficient reads $\mathcal{N}(\lambda_0) = \langle m |\mathcal{Z}(\lambda_0) \mathcal{Z}(\lambda_0)^\dagger | m \rangle^{-1/2}$. Writing the focus power as
\be
P_{\text{foc}}(z, \Delta \lambda) = 
\left< 
\frac{
\vert
\langle m |\mathcal{Z}(\lambda) 
\mathcal{Z}(\lambda_0)^\dagger |m\rangle
\vert^2
}
{ \langle m |\mathcal{Z}(\lambda_0) \mathcal{Z}(\lambda_0)^\dagger | m \rangle}
\right>,
\label{EqPowerFocus}
\ee
 it is clear that it satisfies 
 \be
 P_{\text{foc}}(z, 0) = N P_{\text{rand}}(z) ,
 \label{EqPowerFocusZeroDetuning}
 \ee
 where $P_{\text{rand}}(z) =\left< \langle m |\mathcal{Z}(\lambda_0) \mathcal{Z}(\lambda_0)^\dagger | m \rangle
\right>/N =\left<\vert\mathcal{Z}_{mn} (\lambda_0) \vert^2 \right>$ 
is the fraction of power delivered by random illumination to a speckle grain at depth $z$. Considering $\mathcal{Z}(\lambda_0)$ as a square $N\times N$ matrix, we also have $P_{\text{rand}}(z) = \left< \zeta (z) \right>/N$, where $\left< \zeta (z) \right> = \left<\text{Tr}[\mathcal{Z} (\lambda_0)\mathcal{Z}(\lambda_0)^\dagger ]\right>/N$ is the mean deposition eigenvalue.

At $\Delta \lambda \neq 0$, the mean of the ratio in Eq.~\eqref{EqPowerFocus} can be replaced approximated by the ratio of the means of numerator and denominator,
\begin{align}
P_{\text{foc}}(z, \Delta \lambda) &\simeq \frac{\left<\vert
\langle m |\mathcal{Z}(\lambda) 
\mathcal{Z}(\lambda_0)^\dagger |m\rangle
\vert^2\right>}{\left< \zeta (z) \right>}
\label{EqPowerFocus2}
\\
&= \frac{\sum_{i,j}\left<\mathcal{Z}_{mi}(\lambda)\mathcal{Z}_{mj}(\lambda)^* \mathcal{Z}_{mi}(\lambda_0)\mathcal{Z}_{mj}(\lambda_0)^*\right>}{\left< \zeta (z) \right>}.
\nonumber
\end{align}
This quantity can be computed by decomposing each $\mathcal{Z}$-matrix element on all possible scattering trajectories and performing the average with standard diagrammatic techniques~\cite{Akkermans2007}. We find
\be
P_{\text{foc}}(z, \Delta \lambda) \simeq \left< \zeta (z) \right> \left[\frac{1}{N} + C_1(z,\Delta \lambda )+ C_2(z,\Delta \lambda )\right],
\label{EqPowerFocusMicro}
\ee
where the diagrams corresponding to the short- and long-range contributions $C_1(z,\Delta\lambda)$ and $C_2(z,\Delta\lambda)$ are explicitly represented and evaluated in the supplementary material of Ref.~\cite{McIntosh2024}. The drawback of this approach is that it is difficult to account for all sub-leading diagrams that may become non-negligible after summation over the indices $i,j$. In particular, we note that the expression~\eqref{EqPowerFocusMicro} does not coincide with the exact result~\eqref{EqPowerFocusZeroDetuning} at $\Delta \lambda =0$, since $C_1(z,0) =1$ and $C_2(z,0)>0$. Nevertheless, it meets the anticipated result $P_{\text{foc}}(z, \Delta \lambda \gg \delta \lambda) = P_{\text{rand}} (z)$.  

Alternatively, we can perform a singular value decomposition of each deposition matrix appearing Eq.~\eqref{EqPowerFocus2} in terms of random unitary matrices, $\mathcal{Z}(\lambda)=U\Sigma V^\dagger$, and perform the average in the circular unitary ensemble~\cite{mello90}. This second approach has the clear benefit of properly accounting for all possible contributions. Since Eq.~\eqref{EqPowerFocus2} contains two doublets $\{U, U^\dagger\}$  and two doublets $\{V, V^\dagger\}$, it results in $(2!)^2\times(2!)^2 =16$ terms, of which four dominate in the limit $N\gg1$. The result reads
\begin{align}
P_{\text{foc}}(z, \Delta \lambda) \simeq &\left< \zeta (z) \right> \left[\frac{1}{N} + C_1(z,\Delta \lambda ) \right.
\nonumber
\\
&
\left.+ C_2(z,\Delta \lambda ) + C_V^{(c)}(\Delta \lambda)\right],
\label{EqPowerFocusRMT}
\end{align}
where the $C_1$ and $C_2$ functions are defined in terms of the SVD components, as
\begin{align}
C_1(z,\Delta \lambda) &= \frac{ C_V(\Delta \lambda)C_U(\Delta \lambda)
\left< 
\text{Tr}\left[ \sqrt{\Sigma(z, \lambda)\Sigma(z, \lambda_0)} \right]^2
\right>}{N^2\left< \zeta (z) \right>^2},
\nonumber
\\
C_2(z,\Delta \lambda) &= \frac{C_V(\Delta \lambda)
\left< 
\text{Tr}\left[ \Sigma(z, \lambda)\Sigma(z, \lambda_0) \right]
\right>}{N^2\left< \zeta (z) \right>^2}.
\label{EqC1C2RMT}
\end{align}
Here, $C_V(\Delta \lambda) = N^2\vert \left< V_{i \alpha}(\lambda)  V_{i \alpha}(\lambda_0)^* \right> \vert^2$, and $C_V^{(c)}(\Delta \lambda) = N^2\left< V_{i \alpha}(\lambda)  V_{j \alpha}(\lambda)^*  V_{j \beta}(\lambda_0) V_{i \alpha}(\lambda_0) \right>^{(c)} $ is the connected non-Gaussian part of the correlator made of the product of two doublets $\{V, V^\dagger\}$. The definitions of $C_1(z,\Delta \lambda)$ and $C_2(z,\Delta \lambda)$ in Eq.~\eqref{EqC1C2RMT} coincide with those based on the microscopic diagrammatic picture and introduced in Eq.~\eqref{EqPowerFocusMicro}. In addition, since  $C_V(0)=1$ and $C_V^{(c)}(0)=-1/N$~\cite{mello90}, we find $P_{\text{foc}}(z,0) \simeq N\left[1 + C_2(z,0) \right] P_{\text{rand}}(z)$, which is still different from the exact result~\eqref{EqPowerFocusZeroDetuning} by the extra long-range component $C_2(z,0)$. The reason is not an incorrect evaluation of the numerator of Eq.~\eqref{EqPowerFocus2}, but rather the fact that the approximation~\eqref{EqPowerFocus2} is not valid in the limit $\Delta \lambda \ll \delta \lambda$. To make the result~\eqref{EqPowerFocusRMT}, which is valid for $\Delta \lambda \gtrsim \delta \lambda$, compatible with Eq.~\eqref{EqPowerFocusZeroDetuning}, we propose to replace it by the following empirical expression
 \begin{align}
P_{\text{foc}}(z, \Delta \lambda)&=\left< \zeta (z) \right> \left[  C_1(z,\Delta\lambda) \right.
\nonumber
\\
&+
\left.
\left[1 - C_1(z,\Delta\lambda)\right]
\left[C_2(z,\Delta\lambda) + \frac{1}{N}\right]\right].
\end{align}
This expression is in excellent agreement with the results of numerical simulations at different depths $z$ and arbitrary detuning $\Delta\lambda$, using the numerical values of $C_1(z,\Delta\lambda)$ and $C_2(z,\Delta\lambda)$. The short-range contribution $C_1(z,\Delta\lambda)$ is calculated from the correlation of the field $E(z,\lambda)$ measured in a single speckle and averaged over random input wavefronts,
\begin{align}
    C_1(z,\Delta\lambda)= \frac{\vert\left<E(z,\lambda)E(z,\lambda+\Delta\lambda)^* \right>\vert^2}{\left<E(z,\lambda)E(z,\lambda)^* \right>^2}.
    \label{EqDefC1}
\end{align}
The long-range contribution $C_2(z,\Delta\lambda)$ is calculated from the spectral correlation function of the total intensity $\mathcal{C}^{(T)}(z,\Delta\lambda)$ using the expression, $C_2(z,\Delta\lambda) \simeq \mathcal{C}^{(T)}(z,\Delta\lambda) + C_1(z,\Delta\lambda)/N$, which generalizes the result known for $\Delta\lambda=0$, $\mathcal{C}^{(T)}(z,0) = 2/3N\left< \zeta (z)\right> -1/N$~\cite{Hsu2015}. 

\bibliography{main}
	
\end{document}